\documentclass[conference]{IEEEtran}
\IEEEoverridecommandlockouts
\usepackage{cite}

\usepackage{amsmath,amssymb,amsfonts}
\usepackage{graphicx}
\usepackage{textcomp}
\usepackage{xcolor}
\def\BibTeX{{\rm B\kern-.05em{\sc i\kern-.025em b}\kern-.08em
    T\kern-.1667em\lower.7ex\hbox{E}\kern-.125emX}}

\usepackage[T1]{fontenc}
\usepackage{newtxtext,newtxmath}
\begin{document}

\title{CoNet-Rx: Collaborative Neural Networks for OFDM Receivers}

\author{Mohanad Obeed and Ming Jian\\
Huawei Technologies Canada Co., Ltd., Ottawa, Canada \\
\{mohanad.obeed, ming.jian\}@huawei.com}

\maketitle

\begin{abstract}

Deep learning (DL) based methods for orthogonal frequency division multiplexing (OFDM) radio receivers demonstrated higher signal detection performance compared to the traditional receivers. However, the existing DL-based models, usually adapted from computer vision, aren't well suited for wireless communications. These models require high computational resources and memory, and have significant inference delays, limiting their use in resource-constrained settings. Additionally, reducing network size to ease resource demands often leads to notable performance degradation. This paper introduces collaborative networks (CoNet), a novel neural network (NN) architecture designed for OFDM receivers. CoNet uses multiple small ResNet or CNN subnetworks to simultaneously process signal features from different perspectives like capturing channel correlations and interference patterns. These subnetworks fuse their outputs through interaction operations (e.g., element-wise multiplication), significantly enhancing detection performance. Simulation results show CoNet significantly outperforms traditional architectures like residual networks (ResNets) in bit error rate (BER) and reduces inference delay when both nets have the same size and the same computational complexity.

\end{abstract}


\IEEEpeerreviewmaketitle

\section{Introduction}

The primary components of a communication system are the transmitter, receiver, and channel. Among these, the wireless channel is particularly challenging because it cannot be controlled or precisely predicted. To address this issue, transmitters and receivers are traditionally designed to continuously track channel variations and mitigate their effects. A common technique involves frequently sending pilot signals, enabling the receiver to estimate the channel and subsequently equalize and detect the transmitted symbols. Conventional methods typically utilize least squares (LS) estimation for channel estimation and linear minimum mean square error (LMMSE) for symbol equalization and detection. However, these methods struggle when the wireless channel undergoes rapid fluctuations, as frequent pilot signals alone become insufficient for accurate channel tracking. The main disadvantage of the traditional receivers is that their bit error rate (BER) performance is low in general. The gap between the perfect receiver (that assumes the channels are perfectly known) and the traditional receiver is significant. In addition, they cannot exploit the temporal, spectral, and the spatial correlation to improve the detection performance. Furthermore, in case of interference presence, the traditional receivers mostly deal with the interference as noise, which maximizes their negative impact on the overall performance. 

To overcome the limitations of traditional approaches, deep learning (DL)-based receivers have been introduced, particularly for orthogonal frequency-division multiplexing (OFDM) systems, demonstrating significant performance enhancements over LS-LMMSE methods. DL approaches have shown significant improvements in detecting OFDM received symbols, primarily due to the ability of NNs to extract and leverage temporal, spectral, and spatial correlations to enhance signal detection performance. Additionally, these networks can learn interference patterns and develop strategies to mitigate their effects on detection accuracy. Neural models can replace individual functions in wireless receivers, such as the channel estimator, or even replace a sequence of functions, including channel estimation, equalization, and demapping. To significantly enhance receiver performance, it's crucial to: 1) design a neural network (NN) inputs to maximize the information provided, and 2) develop an efficient network architecture capable of effectively utilizing this information. This paper focuses on designing an efficient NN architecture that leads to a significant increase in the receiver performance. 


Several papers have been proposed deep learning to improve the receivers performance using DL \cite{neumann2018learning, he2018deep, chang2019complex, shental2019machine, zhao2021deep, o2017introduction, dorner2017deep, ait2018end, honkala2021deeprx, honkala2024radio, 11162011}.  Different DL-based approaches are proposed to replace a specific function in wireless receivers. The authors of \cite{neumann2018learning} and \cite{he2018deep} proposed to use NNs for channel estimation. The authors of \cite{chang2019complex} utilized a convolutional neural network (CNN) for equalization, while the authors of \cite{shental2019machine} used a CNN as a demapper that calculates the log-likelihood ratios (LLRs). Those NNs provided better performance compared to the traditional approaches used for channel estimation, equalization, or demapping. Other papers proposed to use NNs to replace multiple functions in the receiver. The authors of \cite{he2018deep} used a multi-layer perceptron NNs to jointly estimate the channels and detect the signals. The authors of \cite{zhao2021deep} used a CNN to directly detect the signals from the received time-domain signals. Other authors (e.g., \cite{o2017introduction, dorner2017deep, ait2018end}) designed end-to-end deep learning based communication system, where the transmitter and the receiver are both trained jointly to form and detect the signals. The authors of \cite{honkala2021deeprx} designed a CNN with residual connections to handle the received frequency-domain OFDM signals and provide the soft bits as outputs. The designed receiver in \cite{honkala2021deeprx} and \cite{honkala2024radio} (called DeepRx) utilizes the frequency and temporal correlations in addition to the pilot signals in every received resource grid to improve the produced soft bits, which leads to minimizing the BER. Several receiver designs proposed in \cite{obeed2025hybrid}, where the traditional LMMSE receiver is integrated with a neural receiver to provide robust detection performance against channel distribution shifts. 

While DL approaches have shown promise in improving the performance of wireless receivers, there are several disadvantages to consider. One of the primary concerns is that most of the NNs employed in radio receivers are either standard CNNs, residual networks (ResNets), or networks adopted from other domains, such as computer vision. While these architectures perform well in many applications, they may not always be optimally suited for the specific needs of wireless communication systems.
Another significant challenge arises when these networks are large and deep, which is typically the case for achieving high performance in complex tasks like signal detection. While deeper networks can provide better accuracy, they come with substantial drawbacks, including large memory requirements, high computational costs, and increased inference delays. These issues are problematic in real-time applications and embedded systems, where resources are limited, and low latency is crucial for optimal performance.
Moreover, reducing the size of these networks, such as by limiting the number of layers or parameters, can lead to a significant drop in the receiver's performance. The trade-off between network complexity and performance is a critical issue in wireless receiver design, as a small reduction in network size may substantially degrade the detection accuracy and the overall reliability of the communication system. Therefore, the need for highly accurate models often conflicts with the practical constraints of memory and processing power, making it difficult to deploy these models in resource-constrained environments, such as mobile or edge devices.

In this paper, we introduce collaborative neural network (CoNet), a novel NN architecture designed for wireless communication receivers, where multiple small ResNets or CNNs sub-networks simultaneously process the received signals' features from diverse perspectives, capturing both channel correlations and interference patterns. These sub-networks collaboratively fuse their outputs through interaction operations (e.g., element-wise multiplications), significantly enhancing detection performance.  Numerical results demonstrate that CoNet outperforms traditional architectures, such as DeepRx \cite{honkala2021deeprx, honkala2024radio}, especially in small-size models, in terms of BER performance under realistic channel conditions. In addition, CoNet can be designed to achieve a substantial reduction in inference delay under a given performance target, thus making it highly suitable for real-time and embedded wireless applications.

The primary contributions of this paper are:
\begin{itemize}

\item We propose CoNet, an innovative NN architecture specifically tailored for wireless receiver applications. Unlike traditional DL models adapted from general-purpose domains, CoNet leverages multiple subnetworks to analyze received signals from distinct perspectives and employs interaction operations (multiplication, addition, concatenation) to effectively fuse these insights.

\item We demonstrate that the proposed architecture significantly improves BER performance compared to traditional architectures (e.g., DeepRx) under realistic simulation scenarios involving varying interference levels, delay spreads, and user mobility.

\item The CoNet architecture achieves comparable performance with significantly fewer layers compared to existing architectures, enabling a substantial reduction in inference latency (approximately 42\%), which is critical for deployment in real-time, resource-limited wireless communication systems.
\end{itemize}

The rest of this paper is organized as follows. The proposed system model is shown in Section II. Section III shows the the proposed CoNet architecture. In Section IV, we present the simulation results. Finally, the paper is concluded in Section V.

\section{System Model}

 We assume an uplink channel, where the user-equipement (UE) is equiped with a single antenna and the receiver is equipped with multiple antennas $N$. Extending this work to multiple-input multiple-output (MIMO) is straight forward. At the transmitter, a sequence of uniformly distributed information bits is randomly generated. These bits undergo encoding using a low-density parity-check (LDPC) code then mapped to symbols, which are allocated across the available physical resource blocks within the transmission time interval (TTI). Demodulation pilots are inserted into designated subcarriers. Subsequently, the data is transformed into an OFDM waveform by applying an inverse fast Fourier transform (IFFT) to the resource blocks, producing $14$ individual OFDM symbols per TTI. To mitigate inter-symbol interference (ISI), a cyclic prefix (CP) is appended to the beginning of each OFDM symbol before transmission. 

 The received signal is assumed to be disturbed by inter-cell interference and noise. The interference is modeled as an additional OFDM waveform that shares the same numerology but carries different random data. This interference signal is transmitted through a separate, randomly generated channel and arrives with a random time offset relative to the target signal. The signal-to-interference ratio (SIR), which defines the power level of the interference compared to the desired signal, is also randomly selected. Therefore, the received signal after fast Fourier transform (FFT) at the $i$th OFDM symbol and the $j$th subcarrier can be expressed as follows
 \begin{equation}
     \mathbf y_{i,j} = \mathbf h_{i,j} x_{i,j} + \mathbf g_{i,j}z_{i,j}+\mathbf n_{i,j},
 \end{equation}
where $\mathbf h_{i,j} \in \mathbb{C}^{N\times 1}$ and $\mathbf g_{i,j} \in \mathbb{C}^{N\times 1}$ are the channels vectors of the transmitter and the source of the interference to the receiver, respectively,   $x_{i,j} \in \mathbb{C}$ and $z_{i,j}  \in \mathbb{C}$ are the transmitted signals from the transmitter and the interferer, respectively, and  $\mathbf n_{i,j}$ is the noise signal.  The powers of $x_{i,j}$ and $z_{i,j}$ are different and selected to obtain a given signal-to-interference and noise ratio (SINR). Given the received signals $\mathbf y_{i,j} \in \mathbb{C}^{N\times 1}$ of all the OFDM symbols and subcarriers in a resource grid (RG), our goal is to design an NN model that finds the corresponding soft bits or the LLRs of the transmitted signals. In other words, the proposed neural receiver replaces the functions of channel estimation, equalization, and demapping. In the following, we describe a common neural receiver (ResNet) proposed in the literature for wireless receivers.

\subsection{ResNets}
\label{Sec:ResNet}

\begin{figure}[t!]
\centering
\includegraphics[scale=0.35]{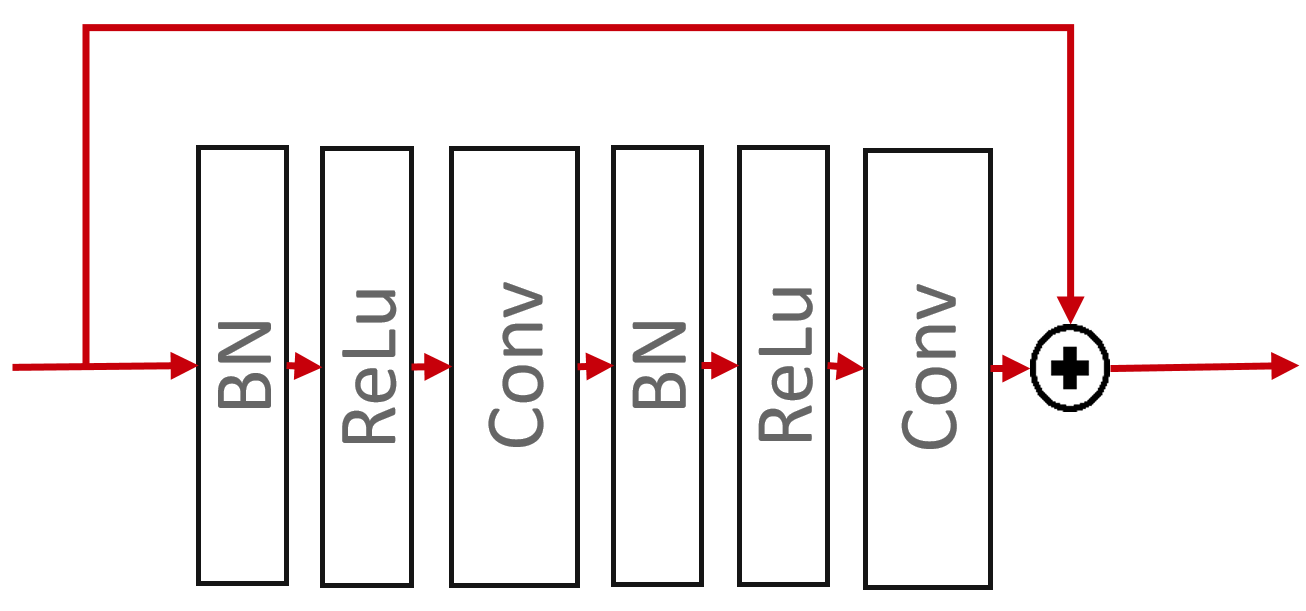}
\caption{Residual block (RB)}
\label{fig:RB}
\end{figure}

The preactivation ResNets introduced in \cite{he2016identity} have been widely applied in wireless receivers \cite{honkala2021deeprx, korpi2021deeprx, luostari2024adapting, lbath2024energy}. Specifically, DeepRx, the neural receiver proposed in \cite{honkala2021deeprx}, demonstrated significant BER performance improvements over traditional receivers. The primary component of these preactivation ResNet architectures is the Residual Block (RB), illustrated in Fig. \ref{fig:RB}. Standard convolutional layers within the RB can alternatively be implemented using depthwise convolutions, consisting of separable convolutions followed by $1 \times 1$ convolutions.

TTI is processed by the ResNet to capture spectral, spatial, and temporal correlations. Thus, the input to the ResNet is represented as a three-dimensional tensor $\mathbf{Y} \in \mathbb{C}^{T \times F \times N}$, where $T$ denotes the number of OFDM symbols per TTI, $F$ indicates the number of subcarriers, and $N$ is the number of receiving antennas. Given that noise and interference variances are known parameters, they are concatenated with the tensor $\mathbf{Y}$ along the third dimension to enhance receiver performance. Furthermore, the pilot symbols are designed to be fixed, allowing the ResNet to efficiently identify and utilize them without requiring dedicated input channels. Since neural networks operate exclusively on real numbers, the complex input tensor $\mathbf{Y}$ is decomposed and concatenated along the third dimension into its real $Re(\mathbf{Y})$ and imaginary $Im(\mathbf{Y})$ parts before being fed into the ResNet.

\section{The Proposed Neural Receiver (CoNet-Rx)}

In this section, we present the proposed neural architecture, CoNet, and how it works as a neural receiver. The CoNet architecture integrates multiple neural networks (NNs) that collaboratively process and interpret received signals from complementary perspectives. Specifically, multiple ResNets or standard CNNs are employed in parallel, each analyzing the received OFDM TTI to extract diverse types of information. This approach enables the networks to collectively enhance BER performance by effectively combining their distinct interpretations of the input signals. For instance, one ResNet might focus on capturing channel correlations across symbols, while another specializes in identifying interference patterns. The resulting information from these subnets is fused through operations such as element-wise multiplication, addition, or concatenation across the layers, ensuring comprehensive feature integration and improved overall performance.

\begin{figure*}[t!]
\centering
\includegraphics[scale=0.48]{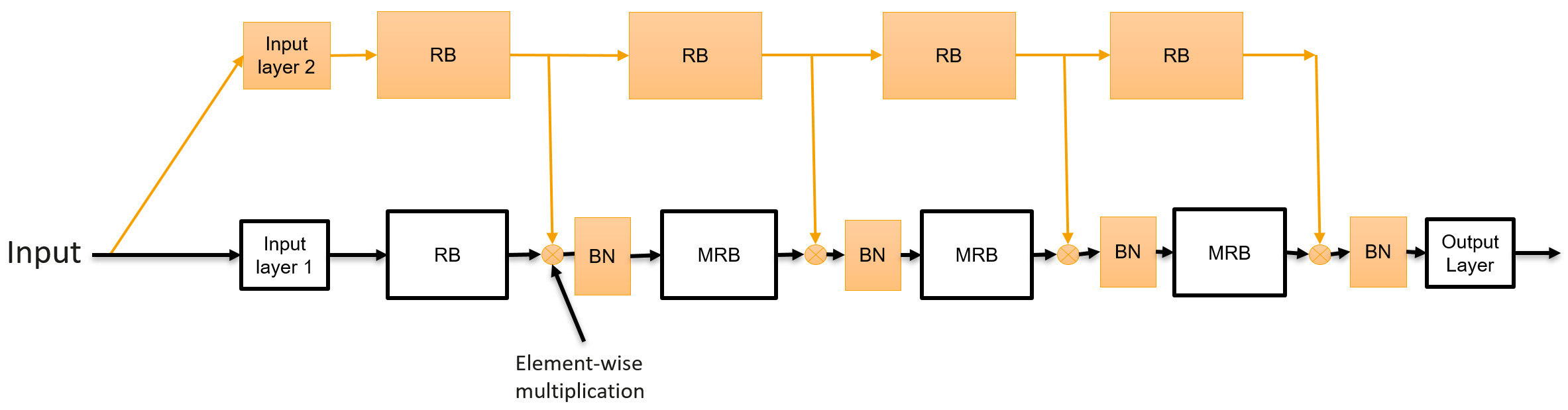}
\caption{Example I of CoNet architecture, RB and MRB blocks are illustrated in figures \ref{fig:RB} and \ref{fig:MRB}, respectively.}
\label{fig:conet}
\end{figure*}

\begin{figure}[t!]
\centering
\includegraphics[scale=0.35]{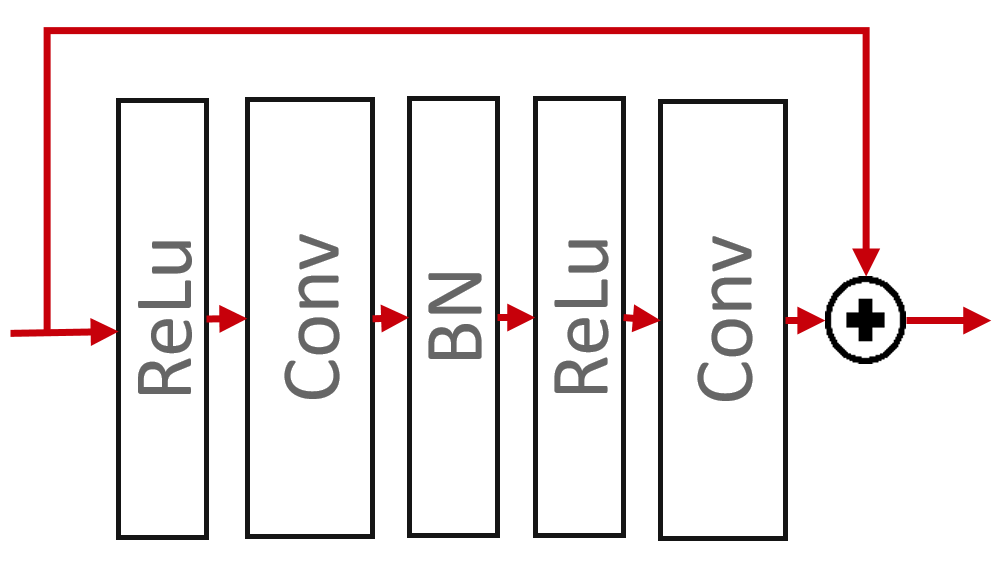}
\caption{Modified residual block (MRB)}
\label{fig:MRB}
\end{figure}

\begin{figure}[t!]
\centering
\includegraphics[scale=0.35]{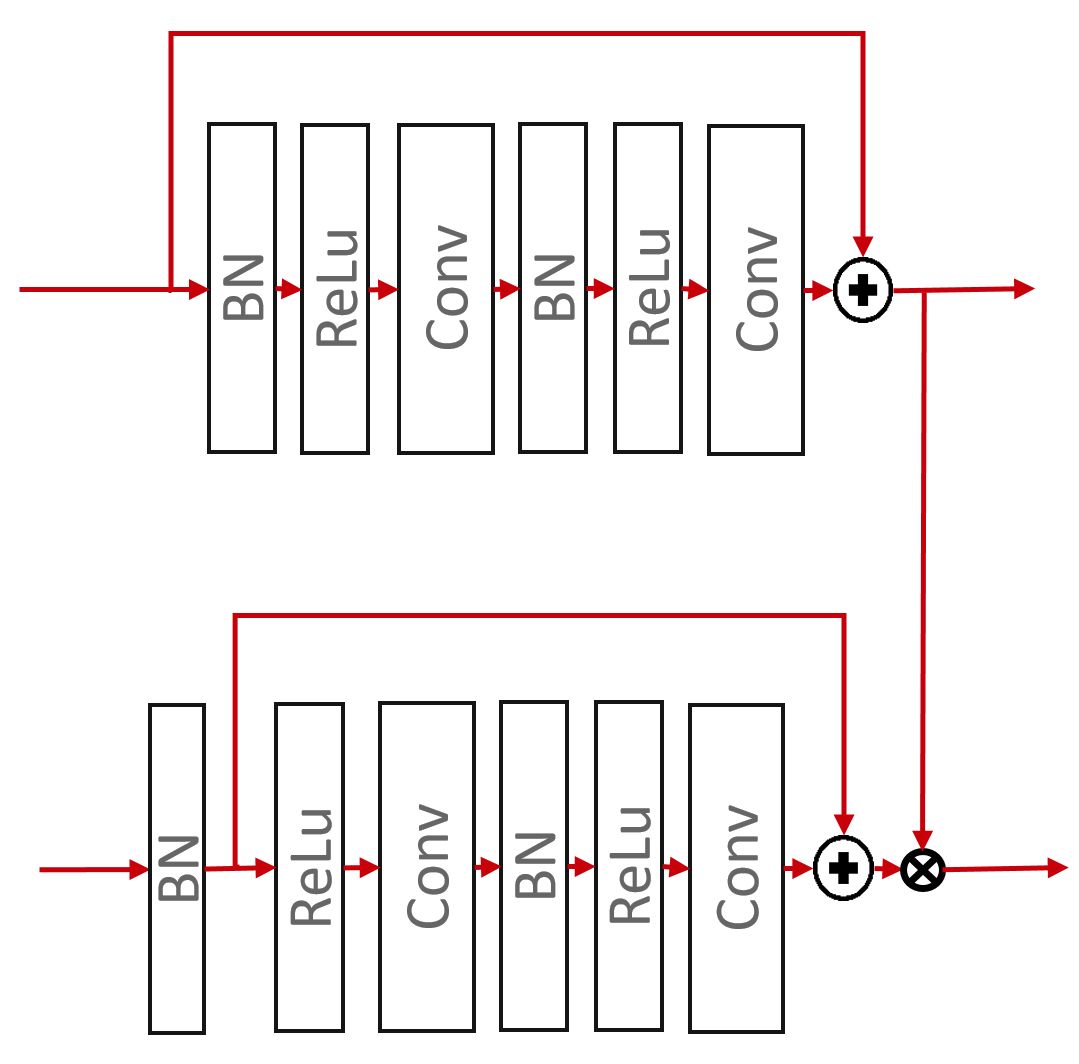}
\caption{Main component of CoNet}
\label{fig:comp}
\end{figure}

Fig. \ref{fig:conet} illustrates an architecture of the proposed CoNet, highlighting two subnets: the main net and the support net that collaborate to minimize the BER. The received TTI, structured as detailed in Section \ref{Sec:ResNet}, is processed concurrently by both subnets. The input layers are convolutional followed by ReLU activations, while the output is a convolutional layer with the number of filters equal to the number of bits per symbol, determined by the modulation order. The output then reshaped to match the required shape at the decoder. Element-wise multiplication is performed between the activation outputs of each subnet at various layers. The resulting activations are then normalized using either batch normalization (BN) or layer normalization (LN) before being fed into subsequent layers of the main net. Incorporating normalization layers immediately following each element-wise multiplication ensures balanced training and mitigates issues such as gradient explosion. Consequently, the activation values flowing through both the main and shortcut paths within subsequent RB/MRB are consistently normalized. This normalization procedure effectively eliminates the necessity for an initial BN in subsequent RBs, modifying the structure of these blocks accordingly, as depicted in Fig. \ref{fig:MRB}. In this architecture, the element-wise multiplication between subnets can be applied uniformly across all RBs outputs or selectively to specific layers, such as the deeper ones. Furthermore, the number of layers or RBs in the support net can vary—being equal to, fewer, or greater than those in the main net. Convolutions within these RBs or modified RBs (MRBs) may employ either standard or depthwise convolutions, providing flexibility and efficiency in network design. The primary component of the proposed CoNet architecture is detailed in Fig. \ref{fig:comp}. Note that each CoNet component consists of two inputs
 and two outputs according to the number of the considered subnets. 



\begin{figure*}[t!]
\centering
\includegraphics[scale=0.38]{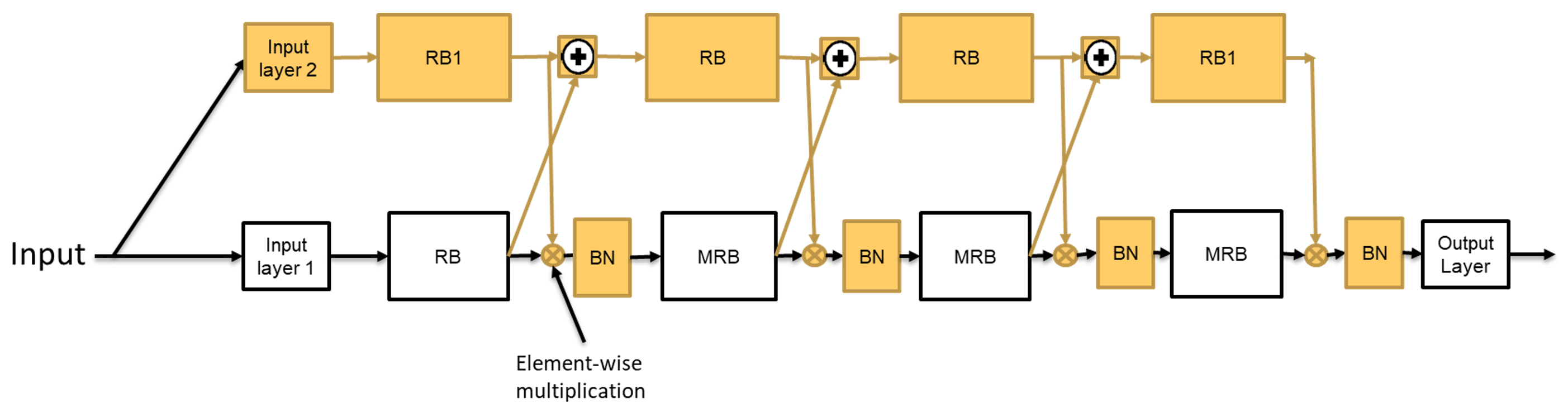}
\caption{Example II of CoNet architecture}
\label{fig:conet2}
\end{figure*}

An alternative CoNet architecture involves bidirectional information flow between subnets. Specifically, the main net can feed information back to the support net across the layers by adding activations from the previous layers of each subnet, as depicted in Fig. \ref{fig:conet2}. This approach demonstrates improved performance, particularly when depthwise convolutions are utilized, and the support net comprises a smaller number of layers.

To generalize the CoNet architectures described earlier, the framework can be extended to incorporate more than two collaborative nets. For instance, one main net could interact with multiple supporting nets ($M$ supporting nets), each providing intermediate information to the main net across layers. Additionally, inputs can be flexibly managed by feeding either the complete TTI into all nets or selectively feeding subsets of the inputs, such as pilot symbols, into specific supporting nets.


It is important to note that employing multiple subnets does not necessarily increase the model size or computational complexity. Model size and computational cost can be effectively managed by adjusting the number of filters in each convolutional layer. For instance, instead of using a single RB with 128 filters per convolutional layer, one can use two parallel RBs with fewer filters per layer, thus maintaining the same total number of parameters and computational complexity. Consequently, to ensure fair comparisons, model sizes are controlled by adjusting filter counts within each layer, resulting in equivalent computational complexity across different CoNet configurations.




\section{Simulation Results}

\begin{figure}[t!]
\centering
\includegraphics[scale=0.45]{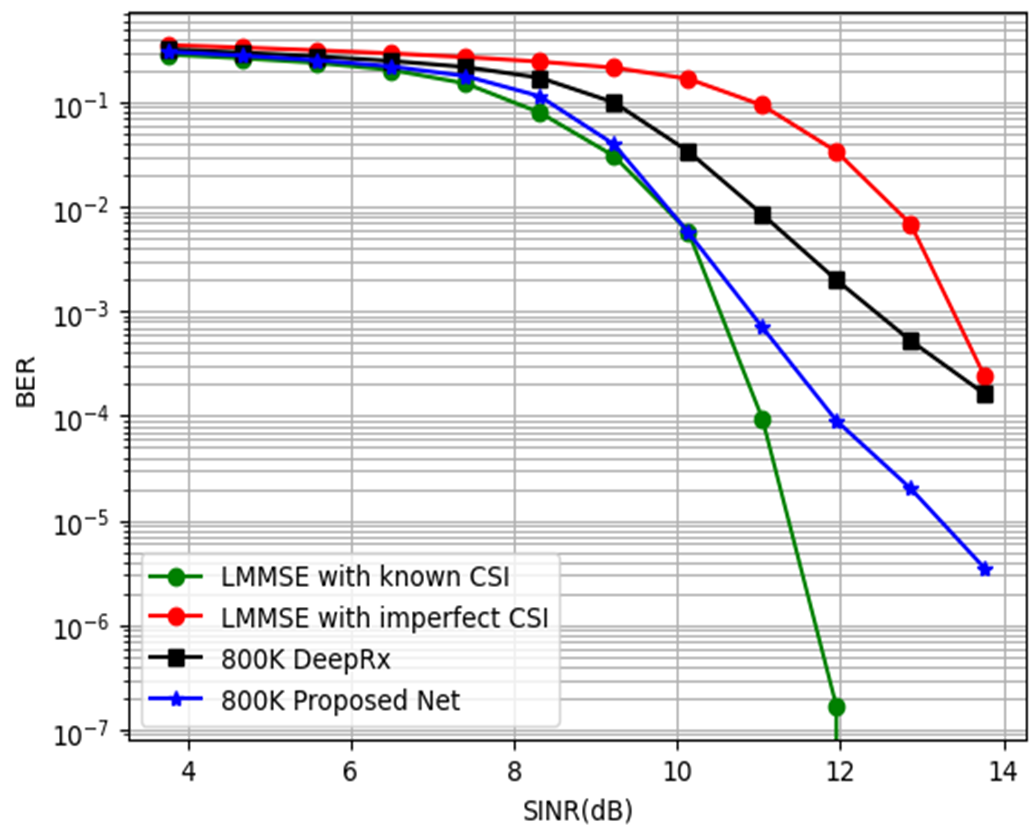}
\caption{BER vs SINR when the interference variance ranging from 10\% to 90\% of the total (noise plus interference) variance.}
\label{fig:Res_0109}
\end{figure}

This section presents the numerical results of the proposed CoNet receiver. We evaluate the BER performance with different values of SINR. We compare the proposed model with preactivation ResNet-based receivers proposed in \cite{honkala2021deeprx}, an LMMSE receiver with imperfect channel-state information (CSI), and an LMMSE receiver with perfectly known CSI. 
The channels are simulated using time-delay-lines (TDL), where the delay spread is uniformly distributed between 1-1500 ns. The user speed is uniformly distributed between 0-15 m/s. Sionna, a Python-based library, has been used to simulate the channels \cite{hoydis2022sionna}. The same dataset is used to train both models. The interference variance used for training is ranging between 10\%-90\% out of the total (noise+interference) given variance. 64-QAM were used as a modulation scheme. In the subsequent results, we exclusively analyze the CoNet model depicted in Fig. \ref{fig:conet}. Although alternative models, such as those shown in Fig. \ref{fig:conet2} or variants employing depthwise convolutions instead of standard convolutions, exhibit comparable performance, their results are omitted here due to space constraints.   Unless stated, the number of parameters in both models (CoNet and DeepRx) is around 800K and the depth of both models is the same (input layer, 4 RBs/MRBs, and an output layer). 

Fig. \ref{fig:Res_0109} compares the performance of the evaluated receiver schemes in terms of BER versus SINR. Specifically, the figure assesses the receivers under varying interference conditions, with interference variance ranging from 10\% to 90\% of the total (noise plus interference) variance. The results clearly demonstrate that the proposed CoNet-Rx substantially outperforms both the DeepRx and the traditional receivers. Notably, CoNet-Rx achieves at least a 1.5 dB improvement over the deepRx for BER values between $10^{-3}$ and $10^{-4}$.

\begin{figure}[t!]
\centering
\includegraphics[scale=0.45]{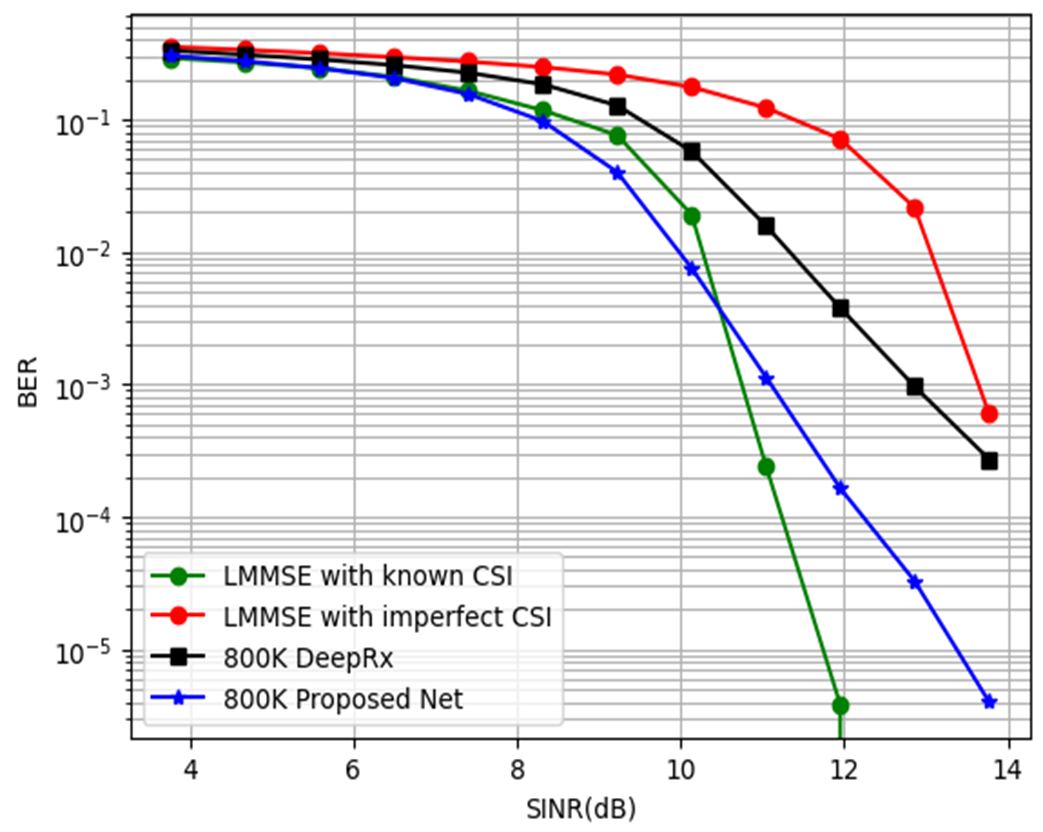}
\caption{BER vs SINR when the interference variance 80\% of the total (noise plus interference) variance.}
\label{fig:highInterf}
\end{figure}
\begin{figure}[t!]
\centering
\includegraphics[scale=0.45]{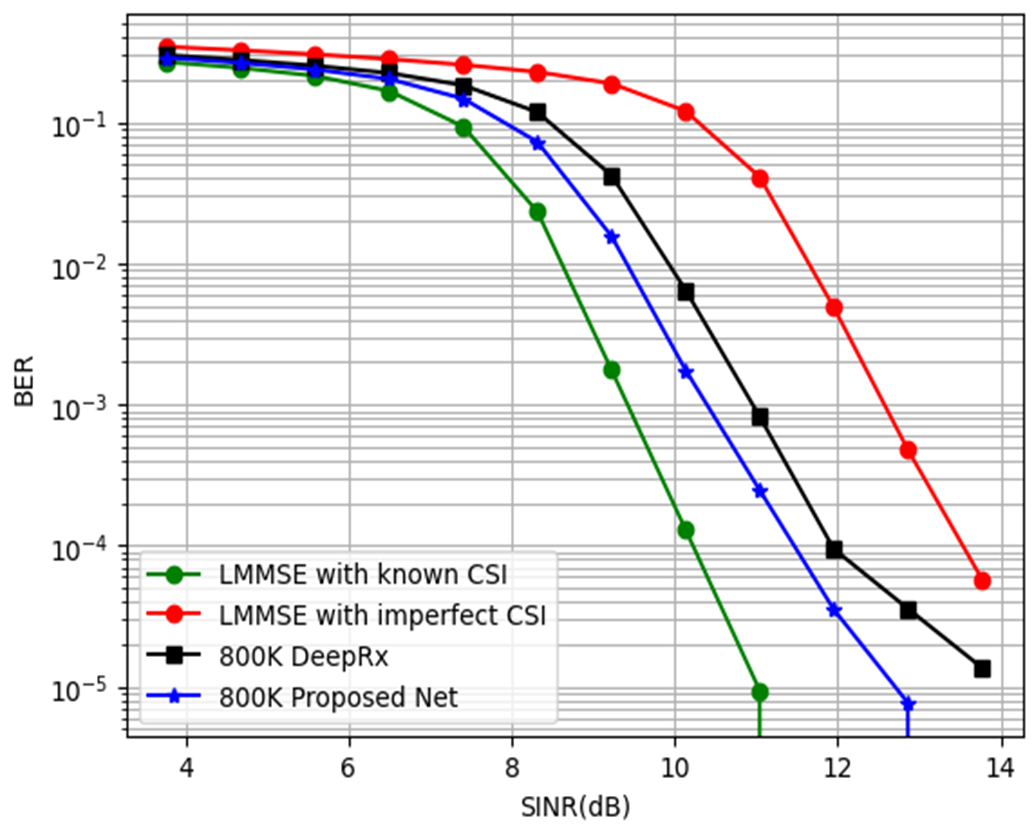}
\caption{BER vs SINR when the interference variance is 30\% of the total (noise plus interference) variance.}
\label{fig:lowInterf}
\end{figure}

Assuming now that the received interference power significantly exceeds the Gaussian noise, Fig. \ref{fig:highInterf} illustrates the performance of the various receivers when interference constitutes 80\% of the total interference and noise variance. The results indicate that the proposed CoNet receiver's performance improves as the interference percentage rises, demonstrating its superior capability to capture and interpret interference patterns compared to standard ResNets. This observation is further supported by the results presented in Fig. \ref{fig:lowInterf}.


Figure \ref{fig:lowInterf} illustrates the performance of the proposed CoNet receiver under low interference levels. When the interference percentage is low (approximately 30\%), CoNet-Rx outperforms other receivers, albeit with a smaller margin compared to scenarios with higher interference levels. This observation suggests that in less challenging environments, where the received signal is relatively clean, the advantage of CoNet's advanced processing capabilities is less pronounced. In such cases, the neural receiver primarily leverages spectral and temporal correlations, with minimal impact from interference patterns.
Conversely, as the complexity of the received signal increases due to higher interference levels or more intricate channel conditions—CoNet's performance advantage becomes more significant. This is attributed to its ability to comprehend and adapt to complex signal patterns, effectively mitigating interference and extracting the desired information. Therefore, CoNet excels in environments where the received signals are more complicated and require nuanced understanding from multiple perspectives.

\begin{figure}[t!]
\centering
\includegraphics[scale=0.45]{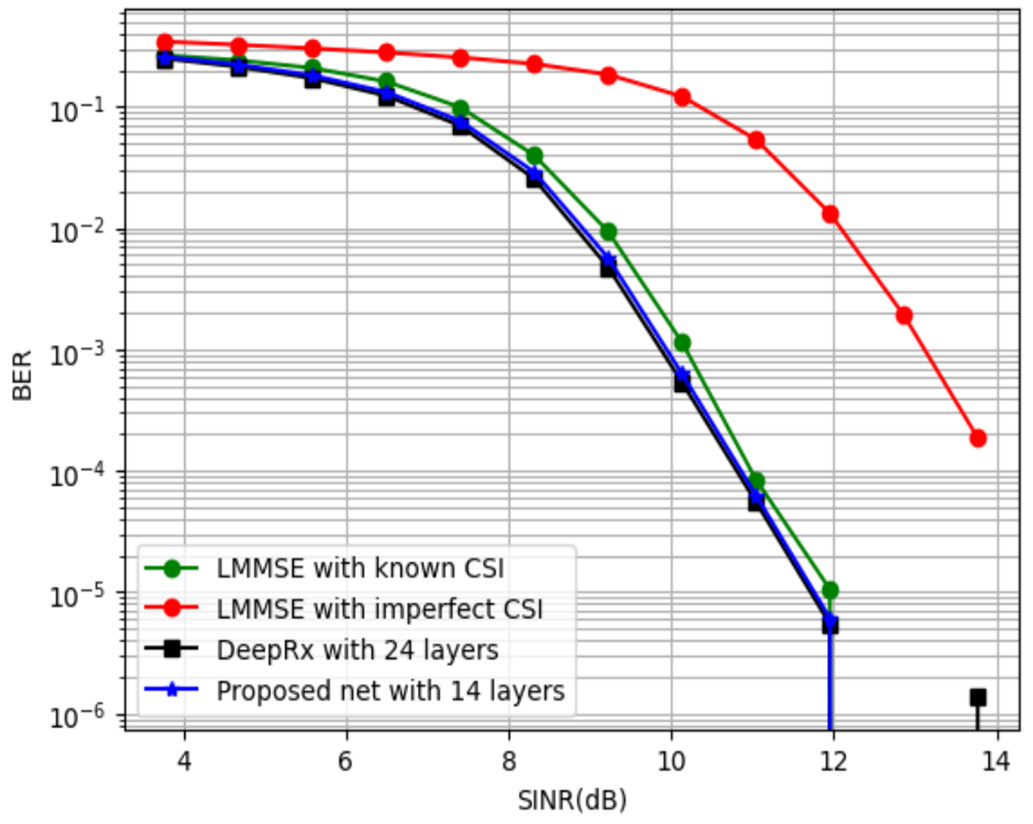}
\caption{BER vs SINR when the number of sequentially-dependent convolutional layers in CoNet and DeepRx are 14 and 24, respectively.}
\label{fig:InferenceD}
\end{figure}

Another advantage of the proposed CoNet architecture is its ability to reduce inference delay when a target performance level is specified. Traditional neural networks, such as DeepRx, often require deeper architectures to achieve high accuracy, which introduces considerable inference latency due to their sequential nature. As shown in Fig. \ref{fig:InferenceD}, the proposed CoNet architecture, with 14 parallel and sequentially dependent convolutional layers, achieves the same performance as DeepRx, which requires 24 sequential layers. Although CoNet is 42\% shallower, both models contain approximately 1.9 million parameters. This parity in parameter count is achieved by carefully selecting the number of filters in each layer. The architectural efficiency of CoNet results in a 42\% reduction in inference time, making it a more suitable choice for latency-sensitive applications.



\section{Conclusion}

This paper introduced CoNet-Rx, a novel neural network architecture specifically designed for OFDM wireless receivers. The proposed CoNet-Rx directly processes frequency-domain signals to detect corresponding soft bits. Distinctively, it comprises multiple collaborating subnets, each analyzing input features from diverse perspectives and interacting across layers through operations such as element-wise multiplications. Simulation results demonstrated that CoNet-Rx significantly outperforms DeepRx, a well-established neural receiver architecture. Notably, while conventional neural receivers suffer considerable performance degradation with reductions in model size and layers, CoNet-Rx maintains high performance even in shallower configurations. Our experiments also showed that a shallow CoNet-Rx achieves performance comparable to a deeper DeepRx model, significantly reducing inference time, a crucial advantage for practical wireless communication systems.


\bibliographystyle{ieeetr}
\bibliography{Ref.bib}

\end{document}